\documentclass[12pt,preprint]{aastex}  
\begin{document} 

\title{Radio and millimeter observations of z~$\sim$~2 luminous QSOs}
\author{A. O. Petric}
\affil{Astronomy Department, Columbia University, New York, NY USA \\
andreea@astro.columbia.edu}
\author{C. L. Carilli}
\affil{National Radio Astronomy Observatory, P.O. Box O, Socorro, NM,
87801, USA \\
ccarilli@nrao.edu}
\author{F. Bertoldi}
\affil{Argelander Institute f\"ur Astronomie, Universit\"at Bonn, Auf dem H\"ugel 71, 5321 Bonn, Germany\\}
\author{A. Beelen}
\affil{Max-Plank-Institut fur Radioastronomie, Auf dem H$\ddot{u}$gel 69, 53121 Bonn, Germany\\}
\author{P. Cox}
\affil{IRAM, 300 rue de la Piscine, Domaine Universitaire, F-38406 Saint Martin d'Heres, France\\}
\author{A. Omont}
\affil{Institute d'Astrophysique de Paris, CNRS, 98bis boulevard Arago, 75014 Paris, France\\}

\begin{abstract}
We present Very Large Array observations at 1.4 and 5 GHz of a sample
of 16 quasi-stellar objects (QSOs) at $z = 1.78$ to 2.71.
Half of the chosen quasars are bright at mm wavelengths (250 or 350~GHz)
while the other half were not detected at mm
wavelengths; the former QSOs were detected
at 1.4~GHz, in most cases at high significance (S/N~$\ge$~7), but only  
three of the latter sources were detected at
radio frequencies, and only at lower significance (S/N~$\sim$~3).
The data are consistent with a correlation  between the mm and radio fluxes
indicating a physical connection between 
the mechanisms responsible for the radio and mm emission. However, this
 conclusion is based on data including many upper limits, and deeper data are clearly
 needed to verify this correlation.
 All eight  mm detected QSOs are detected in the radio
continuum, with  radio flux densities consistent with the
radio-to-FIR correlation for low z star forming galaxies.
However, four of these have flatter spectral indices than
is typical for star forming galaxies (i.e. greater than -0.5) suggesting
that radiation from the central AGN dominates the observed radio
emission. All the sources detected at 1.4 GHz are spatially unresolved, with the
size limits typically $< 1'' = 6$ kpc. High star formation rate
galaxies at low redshift are typically nuclear starbursts, with sizes
$< 1$ kpc. Hence, the current radio size limits are insufficient to
constrain the emission model (AGN or starburst).
\end{abstract}


\section{Introduction}        

Studies of the dynamics of stars and gas in the nuclear regions of
nearby galaxies suggest that the vast majority of spheroidal
galaxies in the nearby universe contain massive black holes and that
the mass of the central black holes correlates with the velocity
dispersion in the spheroid (Magorrian et al. 1998, Ferrarese \&
Merritt 2000, Gebhardt et al. 2000, Tremaine et al. 2002). These findings 
suggest a fundamental relationship between the formation
of massive black holes and the stellar content of galaxies. Radio-millimeter
studies of ($z>2$) QSO's indicate that about 20--30$\%$ of the sources
are detected in surveys with 3 $\sigma$ flux density limits of 1.5 to 4 mJy at 250 GHz
 (Omont et al. 2001, 2003; Carilli et al. 2001a,b; Isaak et al. 2002, 
Bertoldi et al. 2003a,b, Petric et al. 2003, Beelen et al. 2006).  The
mm-to-cm spectral indices of these sources imply that the mm radiation
is thermal emission from warm dust, and that in many of the sources
the spectra are consistent with the dust being heated by star
formation (Carilli et al. 2001b). Detections of CO line emission from
FIR luminous QSOs indicated the presence of large gas reservoirs
($\sim 10^{11}$M$_{\odot}$; Cox et al. 2002, Walter et al. 2003), leading 
some authors to conclude that active star formation is inevitable 
(Omont et al. 2001).

However, the issue of what heats the dust in dusty IR bright QSOs has
not been settled.  Chini, Kreysa, \& Biermann (1989) found that most
{z~$<$~0.4} QSOs show dust emission with dust masses about a few times
{$10^7$~M$_{\odot}$}. On the basis of mm to X-ray spectral indices, these authors
 argue that the dominant dust heating
mechanism is radiation from the AGN. Benford et
al. (1999) analyze a sample of 20 mostly radio quiet quasars with
redshifts between 1.8 and 4.7 by combining their measurements at 350
microns with data from other far-infrared and millimeter
wavelengths. These authors try to fit the spectral energy distribution
and find that the SEDs are consistent with the FIR luminosity being
dominated by emission from a dust component at $\sim 50$K. This is similar to what
is found for ultraluminous infrared galaxies, however for the majority of
sources in their sample a large
luminosity contribution from higher temperature dust cannot be ruled out.
Haas et al. (2000, 2003) obtain infrared to millimeter spectral energy
distributions for a random sample of Palomar Green quasars and
learn that the SEDs show a variety of shapes (power-law, mid-infrared
dominated, and far-infrared dominated). This suggests that more than one type of source 
or heating mechanism or dust are responsible for the obseved SEDs. 

A detailed far-IR photometry study of a sample of optically selected bright quasars
 done by Andreani et
al. (2003) does not find a strong connection between the B luminosity
and the warm dust color temperature and thus suggest that there is no
strong relation between the energy emitted by the nuclear source and
that emitted in the FIR.  McMahon et al. (1994), Isaak et al. (2002), 
Omont et al. (2001), (2003)
find no connection between absolute luminosity and mm/submm flux
(see however, Beelen et al. in preparation for evidence of a mild correlation). 
In particular it should be stressed that the sources of Table 1 are exceptional 
both for their UV and FIR luminosity. Even if the detailed quantitative
correlation between L$_{\rm{FIR}}$ and L$_{\rm{bol}}$ is weak in the luminosity range
considered, the overall probability to have an hyperluminous starburst (L$_{\rm{FIR}}~>~10^{13}$~L$_\odot$)
 in such hyperluminous high z QSOs is high at more than 30\%. Such a correlation
does not necessarily imply that the FIR emission is heated by the AGN since the M$_{BH}$/$\sigma$
Magorrian et al. (1998) relation would also argue for a correlation in objects where the black hole and starburst are
growing coevally. The exceptional FIR luminosity of such QSOs is also consistent with
the rapidly developing 'feedback' models aimed at explaining the M$_{BH}$/$\sigma$
relation (see e.g. Springel et al. 2005, Hopkins et al. 2005 and references therein), as well as
with parallel submm results on X-ray absorbed QSOs (Page et al. 2004) which show a gross
correlation between feeding the QSO and the starburst with interstellar gas. 
	
These results seem to fit in with a picture in which the emission is only loosely
related to the detailed physics of the surrounding medium and so argues in favor of
linking the FIR to a concurrent starburst. Page et al. (2001, see also Page et al. 2004)
 use submillimeter photometry of eight X-ray-absorbed AGN, to argue that the AGN are not sufficiently 
powerful to heat the FIR emitting dust, but that the FIR is
predominantly powered by starlight rather than the central AGN. 
 Similarly, Alexander et al. (2004,2005a,b), observed a population 
of submm SCUBA sources combining ultra-deep
X-ray observations and deep optical spectroscopic data. These authors found that 
a significant fraction ($\geq 38\%$) of the brightest (i.e. 350~GHz flux densities 
greater than 5mJy) SCUBA galaxies do harbor AGNs; but in almost all cases star formation
dominates their bolometric luminosity.

This paper presents the radio (1.4 and 5 GHz) continuum properties of
a sample of {16 QSOs} at z~$\sim$~2. These sources were chosen to (1) have
similar optical properties (M$_B$, spectra) to those of quasar samples 
with redshifts greater than 3.7
 for which we have comparable (sub)millimeter and centimeter
observations and (2) be radio quiet, that is not detected in the Faint
Images of the Radio Sky at Twenty-Centimeter (FIRST) survey which has
a typical sensitivity limit of {$~\sigma ~=~0.15~$mJy} (Becker, White,
\& Helfand 1995).  Half of the observed quasars are bright at 250~GHz
 (S$_{250}$$\ga ~\sim$4~mJy, Omont et al. 2003) or 350~GHz (S$_{350}$$\ga ~\sim$10~mJy,
Priddey et al. 2003) (Table 1) which implies L$_{FIR}$ $>$ 10$^{13}$L$_{\odot}$,  and
 the other half have not been detected at either
of these frequencies. These data help constrain
the evolution with redshift of the radio-to-optical SEDs of luminous
QSOs. We also look for systematic differences between the radio
properties of FIR bright and FIR faint quasars, and examine evidence
for star formation in these systems. We use $\Omega _{\rm{M}}~=~0.3$,
$\Omega _\lambda ~=~0.7$ and H$_0 ~=~71$~km~s$^{-1}$. For comparison
with previous work (e.g. Omont et. al 2003) we also give the 
alternative M$_B$ magnitudes in an Einstein~-~de Sitter (EdS) cosmology, $\Omega _{\rm{M}}~=~1.0$,
$\Omega _\lambda ~=~0$ and H$_0 ~=~50$~km~s$^{-1}$

\section{Observations}

VLA observations at 1.4~GHz were made on April 2002, in the A configuration
(max. baseline = 30 km), with a total bandwidth of 100~MHz and two
orthogonal polarizations. Each source was observed for about one hour
at 1.4~GHz.  Standard phase and amplitude calibration was applied, and
all sources were self-calibrated using field sources. The absolute
flux density scale was set with observations of either 3C48 or 3C286.

The final images were generated using the wide field imaging (Cotton
1999; Bridle \& Schwab 1999) and deconvolution capabilities of the
Astronomical Image Processing System task IMAGR. The theoretical rms
noise ($\sigma$) value corresponding to 1 hour of observing in
continuum mode at 1.4~GHz is {23~$\mu$Jy}, and in most of the maps
presented here this sensitivity is roughly achieved. The Gaussian
restoring CLEAN beam Full Width at Half Maximum (FWHM) was typically
$\sim 1\farcs5$~ for the A configuration observations.

The eight FIR-bright sources were observed at 5~GHz on October, 2002 in
C configuration for $\sim$ 20 minutes, achieving an rms sensitivity of
order 45~$\mu$Jy. The Gaussian restoring CLEAN beam FWHM 
was typically of order 4\arcsec ~for the C configuration observations. The 
lower quality of the 5GHz data did not permit an appropriate fit to determine the source size. 
When determining the spectral indices we used the integrated 5~GHz fluxes and their respective errors.
 
The 250~GHz data, used in the analysis for this
paper, were obtained with the Max-Planck Milimeter Bolometer (MAMBO) array 
(Kreysa et al. 1999) at the IRAM 30m
telescope on Pico Veleta in Spain with 1~$\sigma$ flux density sensitivities
ranging between 0.6 and 1.3~mJy; and are described in detail in Omont
et al. (2003).

\section{Results and Analysis}   

The results for the mm-detected quasars are presented in the upper
half of Table 1, while those for the mm-non-detected sources are
given in the lower half. The 250~GHz data are described by Omont et
al. (2003), while the 350~GHz properties of J0018-0220 and J0035+4405
are detailed by Priddey et al. (2003).

When considering radio emission from the target sources, an important
issue is astrometry and source confusion.  The positional uncertainty
for the radio observations is given by: $\sigma _{\theta} \sim {{FWHM}\over{SNR}}$
  (Fomalont 1999), where FWHM corresponds to that of
the Gaussian restoring beam, and SNR=Signal-to-Noise ratio of the
detection. For a 3$\sigma$ detection this corresponds to 0\farcs5~ for
most of our sources. To this must be added the typical astrometric
uncertainty of the optical data, and the uncertainty in the
relationship between radio and optical reference frame which is about
0\farcs 25 (Deutsch 1999).  Considering confusion,  
Fomalont (2006, in prep.) show that the sub-mJy source counts follow the relation: 
$\rm N(>S_{1.4}) = 0.026 \times S_{1.4}^{-1.1}$ arcmin$^{-2}$, with 1.4 GHz
 flux density,
S$_{1.4}$, in mJy.  Hence, within $1\arcsec .5$ of a given source we expect
$6\times 10^{-4}$ sources with $\rm S_{1.4} \ge 70~\mu$Jy by chance.
A deep survey of 357 square arcminutes by Greve et al. (2004) find that
there are 15 sources in the region surveyed 
with S$_{250} > 3.75$ mJy. At this flux density level we expect about 0.003
sources by chance within the beam of the 30m telescope (i.e. within 5$\arcsec $ of the target source). 
Overall, we only consider radio
emission as associated with the QSO if it is $> 3\sigma $ and located
within 1$\farcs $5~ of its given optical position.

Using these criteria, all of the mm-detected sources are also
detected at 1.4 GHz with flux densities between 0.14 and 0.94 mJy.
Moreover, in all but one case, the sources are detected at high
significance ($> 7\sigma$).  Conversely, only three of the
mm-non-detected sources are detected at 1.4 GHz, and all of these
are fainter than 0.17 mJy. 

The 1.4~GHz images of the detected sources are shown in Figure 1. The
optical positions are indicated by crosses.  From Gaussian fitting we
find no clear evidence that any of the sources are extended at the
1.4~GHz resolution of $\sim~1\farcs5$, with typical upper limits to
source sizes of $\sim 1''$. The 4.8~GHz images of the detected sources
are shown in Figure 2.

\section{The radio-FIR correlation}

In the local universe star-forming galaxies from optically, IR or
radio selected samples follow a very tight linear relation between
their radio continuum and FIR luminosities, with only a factor of two
scatter around linearity over four orders of magnitude in luminosity
(Condon 1992; Condon \& Yin 1990; Miller \& Owen 2001; Yun, Reddy, \&
Condon 2001). If this correlation holds at high redshift (Carilli \&
Yun 1999; 2000; Yun \& Carilli 2002; Elbaz et al. 2002, Appleton et al. 2004)
then the ratio of radio to FIR fluxes can be used to constrain the
star formation properties of high $z$ QSOs. Recent work by Chapman et al. (2005) suggests 
that the radio-FIR correlation also holds in the case of submillimeter
galaxies with redshifts between 1.7 and 3.6. These authors use the 850$\mu$m and radio data with
 the spectroscopic redshift, to estimate dust temperatures for the sources in their sample.
This is done by assuming a spectral energy distribution (SED) like that of an object which would fall
 on the FIR-radio correlation. Chapman et al. (2005) also observe a subset of these galaxies at  450$\mu$m  
to confirm the determined temperatures and SEDs, and so vindicates the assumption that the FIR-radio correlation 
can be used to estimate star formation rates in their sample.

 However, an important
caveat is that lower luminosity, radio quiet QSOs at lower redshift
also follow the standard radio--FIR correlation for star forming
galaxies (Sopp \& Alexander 1991). This can also be seen in the
Haas et al. (2003) sample of PG QSOs. It is unclear whether the sources in the Sopp \&
Alexander sample also host significant active star formation.
Overall, whether or not an object is on the radio-FIR trend is only 
a consistency check that the source is actively forming stars, but certainly 
not conclusive proof thereof. Therefore, it ought to be used in conjunction with
at least one other star formation diagnostic such as radio 
spectral index, variability or source size.

The radio to FIR correlation is typically quantified through the parameter $q$
 defined as:
$$q~= \rm ~log\left({{L_{FIR}}\over{3.75~\times~10^{12}~W \rm{m}^{-2}}}\right)
~-~log\left({F_{1.4}\over{Wm^{-2} Hz^{-1}}}\right)$$
where $L_{\rm{FIR}}$
corresponds to the FIR luminosity  between 40 and 120 $\mu$m and $F_{1.4}$ is
the radio flux at 1.4~GHz in W~m$^{-2}$~Hz$^{-1}$.  In their
extensive study of 2000 IRAS-selected galaxies,  Yun, Reddy \& Condon
(2001) find a mean $q$ value of 2.34, and that 98$\%$ of these
sources are within $\pm$ log(5) of the mean. They conclude that this
range ($q = 1.64$ to 3) corresponds to star forming galaxies, while
significantly lower $q$ values imply a contribution from a radio-loud
AGN.

A second star formation diagnostic is the radio spectral index, with a
typical value of $\alpha \sim -0.75 \pm 0.25$ for star forming
galaxies (Condon 1992). Sub-arcsecond scale radio emission from high
$z$ AGN is typically flat spectrum ($\alpha \sim 0$), although there
are exceptions, namely compact steep spectrum sources (CSOs; O'Dea
1998).  Given the existence of CSOs, we consider the radio spectral
index an additional consistency check for star formation.

Table 2 presents some of the estimated parameters for these sources (names of in Column [1])
which were detected in FIR. Their bolometric luminosities (Column [2]) were estimated
 from the absolute blue magnitudes given by Omont et al. (2003), using
a WMAP cosmology. Omont et al. (2003) used the standard Einstein-de Sitter cosmology 
with {$H_0~=~50~\rm{km~s}^{-1}~\rm{Mpc}^{-1}$,} and $q_0~=~0.5$ to estimate the rest-frame absolute
 B band magnitudes, and so we scaled the blue magnitudes to the concordance cosmology by scaling
the luminosity distances accordingly.  
To convert the blue luminosity $L_{\rm{B}}$ to a bolometric measurement $L_{\rm{bol}}$~we assumed
a bolometric correction from the B band of $L_{\rm{bol}}/L_{\rm{B}}~=~12$ (Elvis et al. 1994).
The radio properties such as the rest frame luminosity at 1.4~GHz  and 5 to 1.4~GHz radio spectral 
indices are shown in columns [3] and [4].
For consistency with earlier treatments (Omont et al. 2001, 2003) the FIR luminosity (column [5]) was derived assuming a dust temperature and dust emissivity of
 45~K 
and an emissivity index $\beta ~=~1.5$ which, in this redshift range, translates to the following scaling: 
$L_{\rm{FIR}}~=~4.7~\times ~10^{12}~({{Dl_{H_0=65}}\over{Dl_{H_0=71}}})^2({{S_{250}}\over{\rm{mJy}}})$~L$_{\odot}$, where ${Dl_{H_0=65}}$~and~${Dl_{H_0=71}}$ are the luminosity distances for the cosmologies used in Omont (2003) and this paper.{\footnote{The FIR properties of these sources are discussed in detail in Omont et al. 2003}} The $q$ 
parameter is given in column [6].
\section{Discussion}
We find a very clear relationship between the mm and radio emission 
properties for the quasars in our sample. Although all the mm-detected sources
 are also detected at 1.4 GHz,
in all cases the 1.4 GHz flux density is much less than the 250 GHz
flux density, typically by an order of magnitude or more. Such a
sharply rising spectrum from the radio through the (rest frame) FIR is
good evidence for the FIR emission being thermal emission from warm
dust (Carilli et al. 2002), as has been confirmed through
multi frequency (rest frame) FIR observations of selected sources,
including some of the sources in the current sample
(Benford et al. 1999; Beelen et al. 2006). 

An important issue concerning the mm fluxes of high $z$
QSOs is whether there is a continuum of mm luminosities, or whether
there are two physically distinct types of QSOs -- mm-loud and
mm-quiet. The flux limits on the non-detections in current mm and submm studies
allow for either possibility. However, Bertoldi et al. (in preparation)
suggest that the average of non-detections yields a clear detection, and that
the distribution of flux densities can be fit by an exponential. On the other hand, 
submm observations of a sample of X-ray absorbed
and non-absorbed AGN find that strong submillimeter emission is found only in 
X-ray absorbed sources (Page et al. 2004, 2001).

 The radio observations presented herein
are interesting in this regard. All eight mm-loud sources in our
study were detected at 1.4~GHz, and all but one of these at high
significance ($\ge 0.2\pm 0.02$ mJy).  Only three of the eight mm-quiet
QSO's were detected at 1.4 GHz, and these at lower flux densities
($\le 0.17$ mJy), and the rest were not detected in images with rms
values $\sim 0.02$ mJy.  Recall that all the sources were selected to
have similar optical magnitudes, such that the radio and mm
differences are unlikely to relate to differences in bolometric
luminosity or magnification by gravitational lensing. We performed a statistical
 study based on methods which take upper limits formally into 
account, using the statistics package ASURV (Fiegelson \& Nelson (1985), Isobe et al. (1986)). 
The statistical tests used indicate that the 
two populations are different to a high significance level, that is the probability that, based on their radio 
properties at 1.4~GHz, the two samples of submm-detected and non-detected
quasars are drawn from the same population is very low ---($\sim 10^{-4}$). 

 Figure 3 shows that a simple explanation for the different 1.4 GHz flux density distributions
  for the submm detected versus non-detected sources may be a correlation between 
the 1.4 GHz and FIR luminosities for QSO host galaxies.
In Figure 3 we show that the 1.4 GHz and FIR luminosities of the sources (including limits), 
are consistent with a correlation between the two bands. This suggests a connection between the mechanisms producing 
the emission in each band. Such a connection would be a natural consequence of star formation. We note that
flux density is not proportional to redshift so this tentative correlation
 cannot result from a mutual correlation with distance. 

However, our samples are very small so it is difficult to ascertain what factors are responsible 
for the lower level of activity in 
the non-FIR detected sources. Clearly, more sensitive surveys of QSOs at multiple  wavelengths are required 
to properly test the hypothesis that there are two physically distinct populations
of mm-loud and mm-quiet sources and to understand the evolutionary stage of these sources. 

As shown in Figure 4, all of the eight mm detected QSOs are detected in the radio
continuum, with  radio flux densities consistent with the
radio-to-FIR correlation for low z star forming galaxies.
However, four of these have flatter spectral indices than
is typical for star forming galaxies ($\alpha ^{5} _{1.4}~>~ -0.5$). 
 
 The other four mm-detected sources are either
at the low end of the $q$ range defined for star forming
galaxies, or have flat radio spectral indices (Table 2). It is
possible that very early in its evolution ($\le 10^6$ years),
ie. before many supernovae have populated the ISM with cosmic ray
electrons, a starburst may show a flat spectral index, corresponding
to free-free radio emission. 
A flat index may be seen if we were observing a very young starburst,
 in which the thermal free-free emission from O and B stars
 overwhelms the synchrotron from SNe. However this scenario would imply that 
we are looking at this object during a rapid and massive starburst.
It seems more likely that in the sources with flat radio indices, the radiation from 
the central AGN dominates the observed radio emission.

J0018-0220, J0812+4028, J1409+5628 and J1611+4719, satisfy both the spectral index and
$q$ parameter criteria for star forming galaxies, although we
re-emphasize neither test is definitive in this regard. 
If the dust is heated by star formation, the implied star formation
rates are of order $10^3$ M$_\odot$ year$^{-1}$ (Omont et al. 2003).
At this rate, assuming a Salpeter IMF, a large fraction of the stars in the host spheroidal
galaxy could form on a typical dynamical timescale of $10^7$ to $10^8$
years.  For one source, J1409+5628, a massive reservoir of molecular
gas ($\sim 10^{11}$ M$_\odot$), the required fuel for such a
starburst, has recently been detected (Beelen et al. 2004). Similar
searches are underway for CO emission from the other QSOs.

All the sources detected at 1.4 GHz are spatially unresolved, with the
size limits typically $< 1'' = 6$ kpc. High star formation rate
galaxies at low redshift are typically nuclear starbursts, with sizes
$< 1$ kpc. Hence, the current radio size limits are insufficient to
constrain the emission model (AGN or starburst). A potentially
powerful test of AGN vs. starburst radio emission comes from VLBI
observations, to search for extended (on scales of 100's of pc), lower
surface brightness radio emission.  Recent VLBI observations 
(Momjian, Petric \& Carilli 2004, Momjian, Carilli, \& Petric 2005) of one of
the sources in our sample (J1409+5628), show extended emission on
scales of 100 pc with an intrinsic brightness temperatures $< 10^5$ K
at 8 GHz, as expected for a starburst nucleus (Condon \& Yin 1990),
and inconsistent with a radio-jet source (CSO or core-jet; Beelen et
al. 2004).

\vskip 0.2in
CC would like to acknowledge support from the Max-Planck-Forschungspreis.
The National Radio Astronomy Observatory (NRAO) is
a facility of the National Science Foundation, operated under
cooperative agreement by Associated Universities, Inc.
We also thank an anonymous referee that helped improve the structure and 
content of this paper.

\clearpage\newpage
\thispagestyle{empty}
\begin{deluxetable}{lccccccccc}
\rotate
\tableheadfrac{0.0}
\tablecolumns{10}  
\tabletypesize{\scriptsize}    
\tablewidth{0in}
 \tablecaption{Observed Properties}
\tabletypesize{\scriptsize}
\tablehead{
\colhead{}
&\colhead{}
&\colhead{}
&\multicolumn{2}{c}{Optical Position (J2000)}
&\multicolumn{2}{c}{Radio 1.4~GHz Position (J2000)}
&\colhead{S$_{1.4}$}
&\colhead{S$_{5.0}$}
&\colhead{S$_{250}$}\\
\colhead{QSO}
&\colhead{$z$}
&\colhead{M$_{\rm{B}}^{\lambda}$~(M$_{\rm{B}}~^{\rm{EdS}}$)}
&\colhead{RA ($h~m~d$)} &\colhead{DEC (\arcdeg ~ \arcmin ~\arcsec~)}
&\colhead{RA ($h~m~d$)} &\colhead{DEC (\arcdeg ~ \arcmin ~\arcsec~)}
&\colhead{[1.0e-6~Jy]}
&\colhead{[1.0e-6~Jy]}
&\colhead{[1.0e-3~Jy]}\\
}
\startdata
\cutinhead{Properties of sources detected in mm/mm}
J0018-0220&2.56 &-28.4(-28.3)&00 21 27.37 &-02 03 33.8 & 00 21 27.24&-02 03 33.6&260~$\pm$~20&$<$120&$**$\tablenotemark{a}\\
J0035+4405&2.71&-28.5(-28.4)&00 37 52.31 & 44 21 32.9 & 00 37 52.33&44 21 32.9&150~$\pm$~16&$<$~120~$$&$**$\tablenotemark{b}\\
J0812+4028&1.78&-27.0(-27.0)&08 12 00.50 & 40 28 14.0 & 08 12 00.50&40 28 14.3&200~$\pm$~60&$<~100$&4.3$\pm$~0.8\tablenotemark{c}\\
J0937+7301&2.52&-28.6(-28.5)&09 37 48.89 & 73 01 58.3\tablenotemark{d} & 09 37 48.89&73 01 58.1&440~$\pm$~20&530~$\pm$~36&3.8$\pm$~0.9\tablenotemark{c}\\
J1409+5628&2.56&-28.5(-28.4)&14 09 55.60 & 56 28 26.2\tablenotemark{d} & 14 09 55.57&56 28 26.5&940~$\pm$~20&310~$\pm$~60&10.7$\pm$~0.6\tablenotemark{c}\\
J1543+5359&2.37&-28.4(-28.3)&15 43 59.37 & 53 59 03.1\tablenotemark{d} & 15 43 59.45&53 59 03.3&140~$\pm$~20&370~$\pm$~100&3.8$\pm$~0.9\tablenotemark{c}\\
J1611+4719&2.35&-27.8(-27.7)&16 12 39.90 & 47 11 58.0 &16 12 39.91&47 11 57.6& 200~$\pm$~20&$<$~120&4.6$\pm$~7\tablenotemark{c}\\
J1649+5303&2.26&-28.2(-28.2)&16 49 15.02 & 53 03 16.5\tablenotemark{d} &16 49 15.00&53 03 16.6& 820~$\pm$~20&910~$\pm$~80&4.6$\pm$~0.8\tablenotemark{c}\\
\cutinhead{Properties of sources with mm/mm upper limits}
J0837+145&2.51&-28.2(-28.1)&08 37 12.60&14 59 17.0&$...$&$...$&$<$81&$...$&$<$~1.8\tablenotemark{c}\\
J0958+470&2.48&-27.8(-27.7)&09 58 45.50&47 03 24.0&$...$&$...$&$<$81&$...$&$<$~2.1\tablenotemark{c}\\
J0914+582&1.95&-27.1(-27.1)&09 14 25.80&58 25 19.0&09 14 25.74 &58 25 19.4&120~$\pm$~31&$...$&$<$~1.8\tablenotemark{c}\\
J1210+3939&2.40&-27.8(-27.7)&12 10 10.20&39 39 36.0&12 10 10.22&39 39 35.7&74~$\pm$~ 24&$...$&$<$~2.4\tablenotemark{c}\\
J1304+2953&2.85&-28.1(-28.0)&13 04 12.00&29 53 49.0&13 04 11.97&29 53 49.1&88~$\pm$~ 22&$...$&$<$~3.0\tablenotemark{c}\\
J1309+2814&2.21&-27.9(-27.9)&13 09 17.20&28 14 04.0&$...$&$...$&$<$66&$...$&$<$~3.0\tablenotemark{c}\\
J1401+5438&2.37&-27.4(-27.3)&14 01 48.40&54 38 59.0&$...$&$...$&$<$72&$...$&$<$~2.7\tablenotemark{c}\\
J1837+5105&1.98&-29.3(-29.3)& 18 37 25.30& 51 05 59.0&$...$&$...$&$<$60&$...$&$<$~2.4\tablenotemark{c}\\
\enddata
\tablenotetext{a}{Priddey et al. 2002 detect this source at 350~GHz at 17.2$\pm$~2.9~mJy. At the time of
 submitting this paper no observations at 250GHz were made of this source. }
\tablenotetext{b}{Priddey et al. 2002 detect this source at 350~GHz at 9.4$\pm$~2.8~mJy. At the time of
 submitting this paper no observations at 250GHz were made of this source.}
\tablenotetext{c}{Omont et al. 2003}
\tablenotetext{d}{Position derived from the Digital Sky Survey using AIPS program JMFIT}
\tablecomments{The source name is given in column [1], its
redshift in [2], absolute blue magnitude (M$_{\rm{B}}$) in [3], its
optical position in J2000 coordinates in [4], and [5]. The location of
the 1.4~GHz radio emission peaks as derived from the Gaussian fitting
is given in columns [6] and [7]. The flux densities at 1.4, 5 and 250
are given in in columns [8], [9], and [10] respectively with 1$\sigma$
error bars. In cases of non-detections, 3~$\sigma$ upper limits to
flux densities are listed..}
\end{deluxetable}

\clearpage\newpage
\pagestyle{empty} 
\begin{deluxetable}{lcccccc}
\tableheadfrac{0.0}
\tablecolumns{7}  
\tabletypesize{\scriptsize}    
\tablewidth{0in}
 \tablecaption{Estimated Properties}
\tabletypesize{\small}
\tablehead{
\colhead{QSO}
&\colhead{L$_{\rm{bol}}$}
&\colhead{${{\alpha} ^{5}} _{1.4}$}
&\colhead{L$_{1.4}$}
&\colhead{L$_{5.0}$}
&\colhead{L$_{\rm{FIR}}$}
&\colhead{$q$}\\
\colhead{}
&\colhead{L$_{\odot}$}
&\colhead{}
&\colhead{W Hz$^{-1}$}  
&\colhead{W Hz$^{-1}$}
&\colhead{L$_{\odot}$}
&\colhead{}\\
}
\startdata
J0018-0220 &2.20e+14 &$\le$~-0.61& 8.3e+24&3.8e+24&2.3e+13&2.4~$\pm$~0.2\\
J0035+4405 &2.42e+14 &$\le$~-0.18& 3.1e+24&2.5e+24&1.3e+13&2.3~$\pm$~0.3\\
J0812+4028 &6.07e+13 &$\le$~-0.60& 2.7e+24&1.4e+24&1.7e+13&2.7~$\pm$~0.4\\
J0937+7301 &2.65e+14 & 0.15~$\pm$~0.08&5.3e+24&6.4e+24& 1.5e+13&2.4~$\pm$~0.2\\
J1409+5628 &2.42e+14 &-0.90~$\pm$~0.25&4.2e+25&1.4e+25& 4.3e+13&2.0~$\pm$~0.1\\
J1543+5359 &2.20e+14 & 0.80~$\pm$~0.20&7.2e+23&1.9e+24& 1.6e+13&3.3~$\pm$~0.3\\
J1611+4719 &1.27e+14 &$\le$~-0.40&4.2e+24&2.5e+24& 1.9e+13&2.5~$\pm$~0.2\\
J1649+5303 &1.83e+14 & 0.10~$\pm$~0.30&8.9e+24&9.9e+24& 1.9e+13&2.4~$\pm$~0.2\\
J0837+145  &1.83e+14 &$...$&$\le$~3.0e+24&$\le$~1.1e+24&$\le$~7.3e+12&$...$\\
J0958+470  &1.27e+14 &$...$&$\le$~2.9e+24&$\le$~1.1e+24&$\le$~8.5e+12&$...$\\
J0914+582  &6.66e+13 &$...$&2.9e+24&9.6e+23&$\le$~7.3e+12&$\le$~2.5\\
J1210+3939 &1.27e+14 &$...$&2.5e+24&9.5e+23&$\le$~1.2e+13&$\le$~2.6\\
J1304+2953 &1.67e+14 &$...$&4.3e+24&1.7e+24&$\le$~1.2e+13&$\le$~2.5\\
J1309+2814 &1.39e+14 &$...$&$\le$~1.8e+24&$\le$~7.0e+23&$\le$~1.2e+13&$...$\\
J1401+5438 &8.77e+13 &$...$&$\le$~2.3e+24&$\le$~8.9e+23&$\le$~1.1e+13&$...$\\
J1837+5105 &5.05e+14 &$...$&$\le$~1.3e+24&$\le$~5.0e+23&$\le$~9.7e+13&$...$\\
\enddata
\tablecomments{Table 2 presents some of the estimated properties for the
mm-detected sources (Column [1]). Their bolometric luminosities
(Column [2]) were estimated from the absolute blue magnitudes given in
Omont et al. (2003), using a WMAP cosmology. Omont et al. 
used the standard Einstein-de Sitter cosmology with
{$H_0~=~50~\rm{km~s}^{-1}~\rm{Mpc}^{-1}$,} and $q_0~=~0.5$ to estimate
the rest-frame absolute B band magnitudes, and so we scaled the blue
magnitudes to the concordance cosmology by scaling the luminosity distances.
 To convert the blue luminosity $L_{\rm{B}}$ to a bolometric measurement $L_{\rm{bol}}$~we
assumed a bolometric correction from the B band of
$L_{\rm{bol}}/L_{\rm{B}}~=~12$ (Elvis et al. 1994). The radio
properties (1.4~GHz luminosity and 5 to 1.4~GHz spectral index,
$\alpha_{1.4}^5$), the 1.4 and 5~GHz rest-frame luminosities
 are gvine in columns [3], [4] and [5]. For the objects observed at both 1.4 and 5 GHz we
 calculated the rest-frame 1.4~GHz luminosity using the estimated spectral indices, or the upper limits for the indices. 
For sources observed at only one radio frequency (1.4~GHz) we used a default spectral index of -0.75. 
The FIR luminosity (column [6]) was derived assuming a dust temperature of 45~K, and an emissivity index $\beta ~=~1.5$. This
translates to the following scaling: $L_{\rm{FIR}}~\sim~4.7~\times
~10^{12}~({{S_{250}}\over{\rm{mJy}}})$~L$_{\odot}$ over the $z$ range
of interest. The $q$ parameter is given in column [7].}
\end{deluxetable} 
\clearpage\newpage
\centerline{Figure Captions}
{\bf Figure 1:} Images at 1.4~GHz of the sources detected at 1.4~GHz discussed in this paper.
The FWHM of the Gaussian restoring beams are shown in the insets in all frames. 
Contour levels (solid lines) are a geometric progression in the square root of two 
starting at 2$\sigma$, with $\sigma$ listed below ($\sigma$ corresponds
to the measured rms on the image). Three negative contours starting at -2$\sigma$ ~(dashed) are included. 
 The central cross in each image marks the optical QSO location.  \\
 Fig1a = J0018-0220, $\sigma$ = 17 $\mu$Jy beam$^{-1}$; 
 Fig1b = J0035+4405, $\sigma$ = 16 $\mu$Jy beam$^{-1}$;
 Fig1c = J0812+4028, $\sigma$ = 27 $\mu$Jy beam$^{-1}$;
 Fig1d = J0937+7301, $\sigma$ = 21 $\mu$Jy beam$^{-1}$;
 Fig1e = J1409+5628, $\sigma$ = 23 $\mu$Jy beam$^{-1}$;
 Fig1f = J1543+5359, $\sigma$ = 19 $\mu$Jy beam$^{-1}$;
 Fig1g = J1611+4719, $\sigma$ = 19 $\mu$Jy beam$^{-1}$	;
 Fig1h = J1649+5303, $\sigma$ = 19 $\mu$Jy beam$^{-1}$;
 Fig1i = J0914+582, $\sigma$ = 30 $\mu$Jy beam$^{-1}$;
 Fig1j = J1210+3939, $\sigma$ = 25 $\mu$Jy beam$^{-1}$;
 Fig1k = J1304+2953, $\sigma$ = 22 $\mu$Jy beam$^{-1}$;

{\bf Figure 2:} Images at 4.8~GHz of the sources detected at this frequency.
The FWHM of the Gaussian restoring beams are shown in the insets in all frames. 
Contour levels (solid lines) are a geometric progression in the square root of two 
starting at 2$\sigma$, with $\sigma$ listed below ($\sigma$ corresponds
to the measured rms on the image). Two negative contours (dashed) are included. 
 The central cross in each image marks the optical QSO location.  \\
 Fig1d = J0937+7301, $\sigma$ =  $\mu$Jy beam$^{-1}$;
 Fig1e = J1409+5628, $\sigma$ =  $\mu$Jy beam$^{-1}$;
 Fig1f = J1543+5359, $\sigma$ =  $\mu$Jy beam$^{-1}$;
 Fig1h = J1649+5303, $\sigma$ =  $\mu$Jy beam$^{-1}$;

{\bf Figure 3:} Radio(in W Hz$^{-1}$) versus mm luminosities (in L$_{\odot}$ ) for all our sources. The doted line represents the relation between 1.4~GHz and FIR luminosity for a star-forming object. This figure 
suggests that the mechanisms producing radio and mm emission in these quasars are connected. We note that
flux density is not proportional with redshift so this tentative correlation
 cannot result from a mutual correlation with distance. 

{\bf Figure 4:} Distribution of $q$ values plotted as a function of 60~$\mu$m
luminosity.  The crosses are for the IRAS 2Jy sample of Yun et al. (2001). 
The solid squares with error bars are for the mm-detected QSOs with steep
radio spectral indices and the empty triangles are mm and radio detected
source with almost flat or raising spectral indices (Table 2). 
The solid line marks the average value of $q ~= ~2.34$,
while the dotted lines mark the radio-excess (below) and IR-excess
(above) objects, as discussed in Yun et al. (2001).

 \pagestyle{empty} 
\clearpage\newpage
\begin{figure}
{{\plottwo{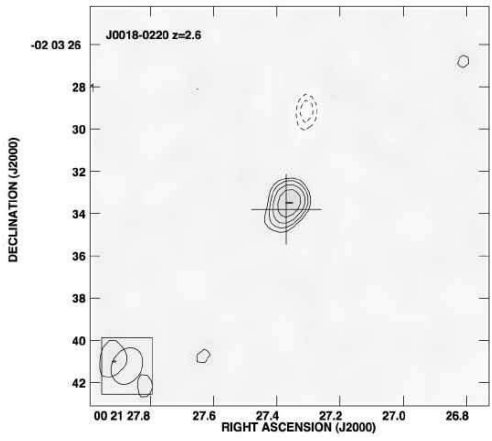}{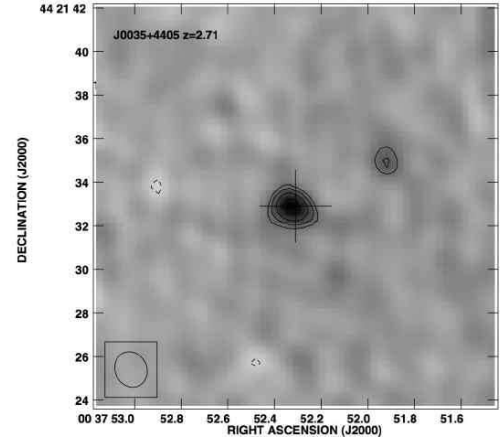}}
 {\plottwo{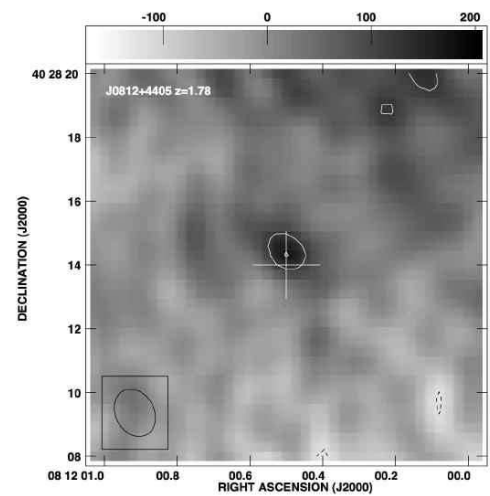}{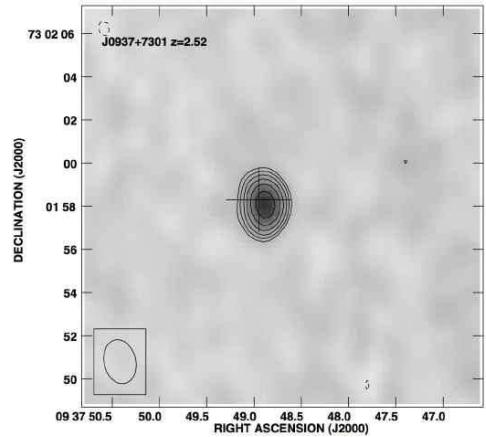}}}
\caption{Images at 1.4 GHz}
\end{figure}
\clearpage\newpage
\begin{figure}
{{\plottwo{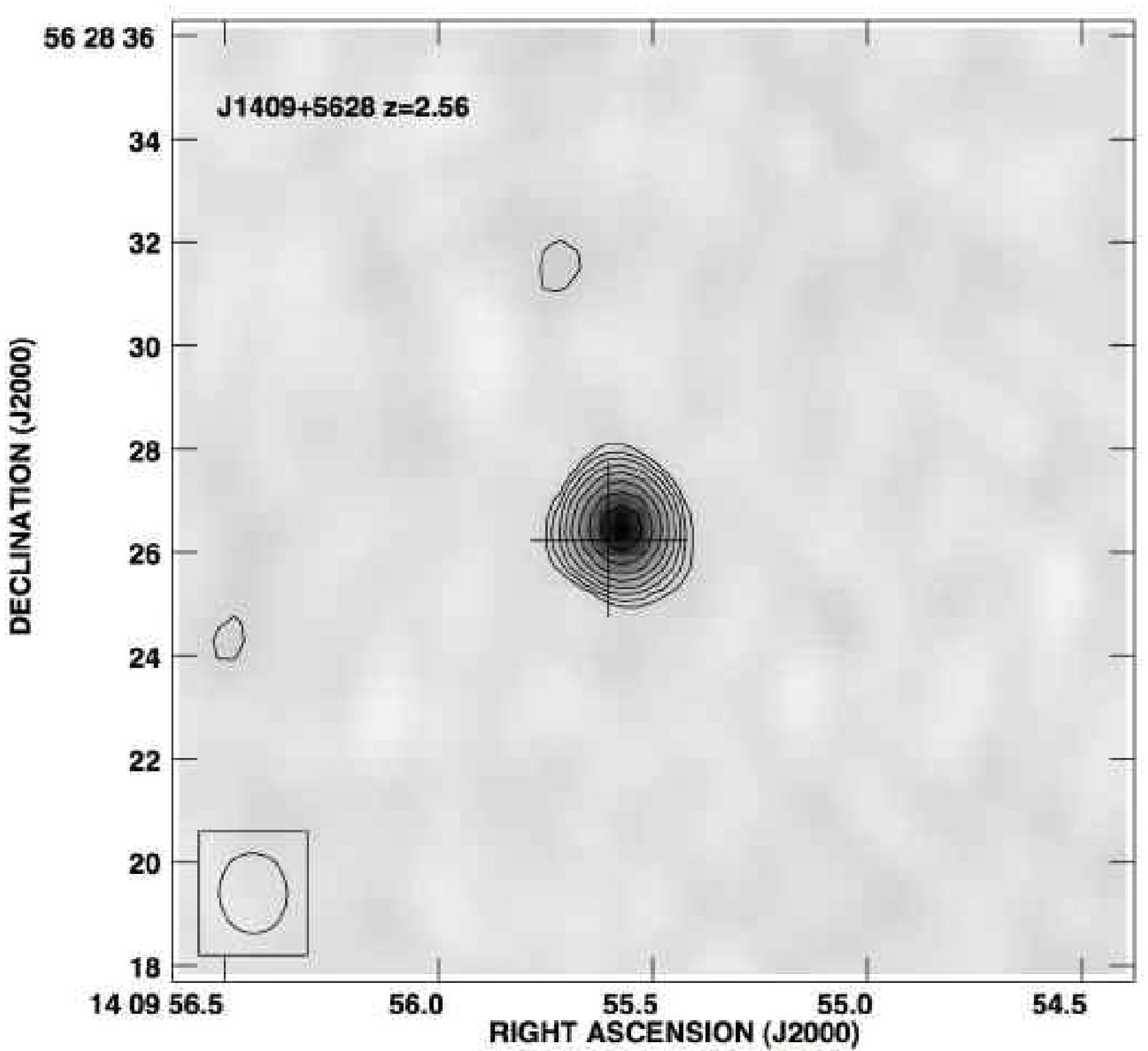}{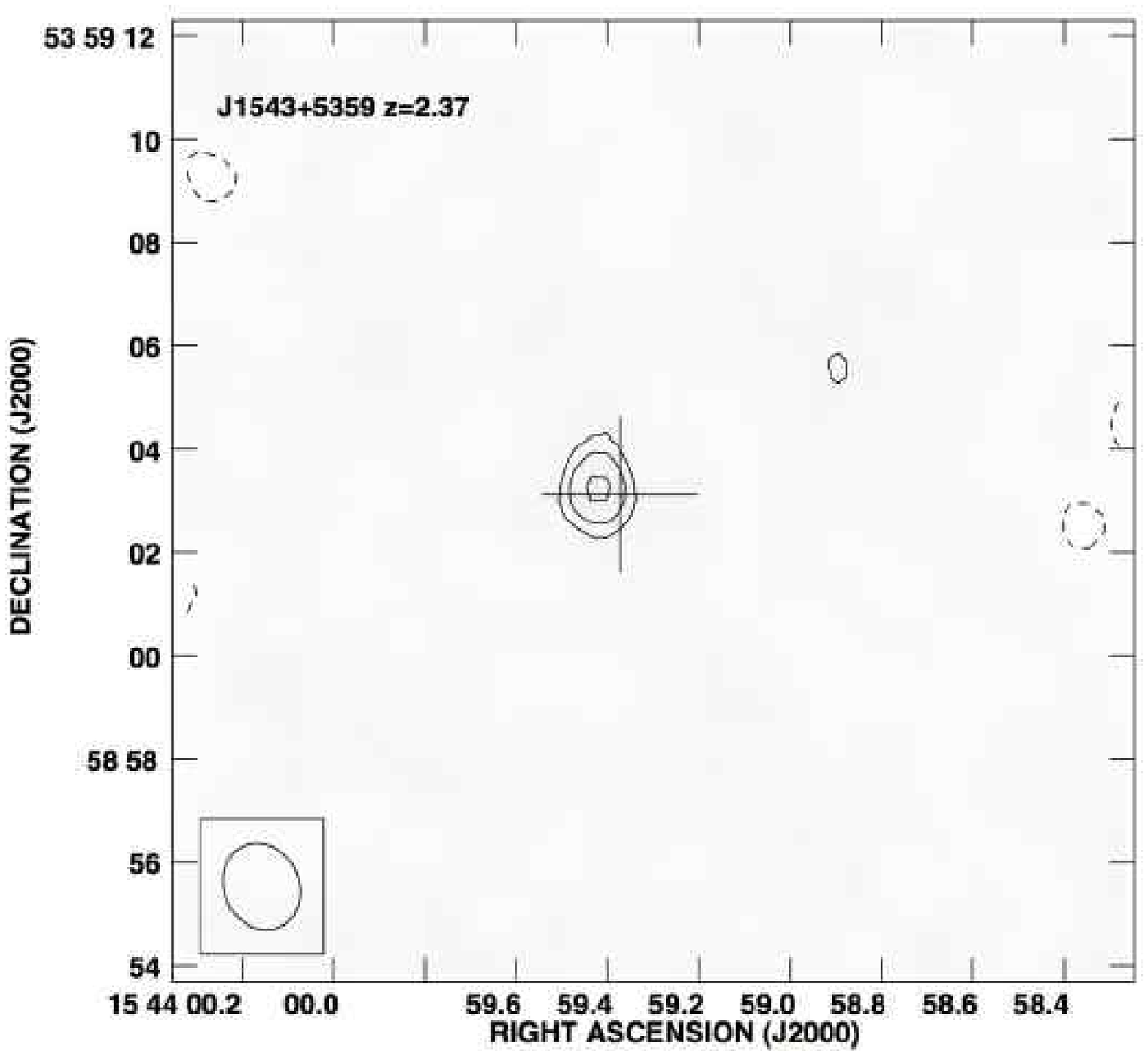}}
 {\plottwo{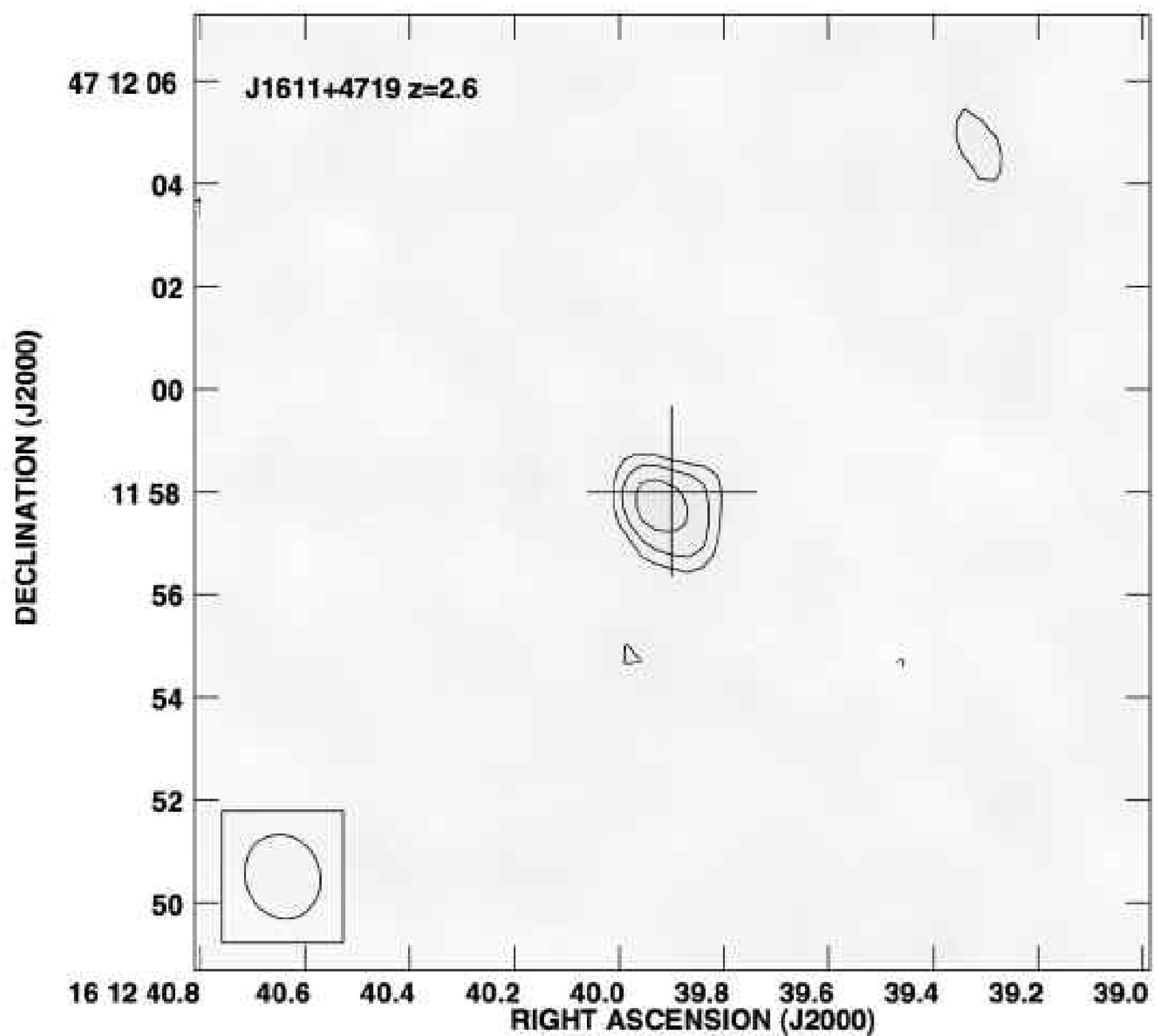}{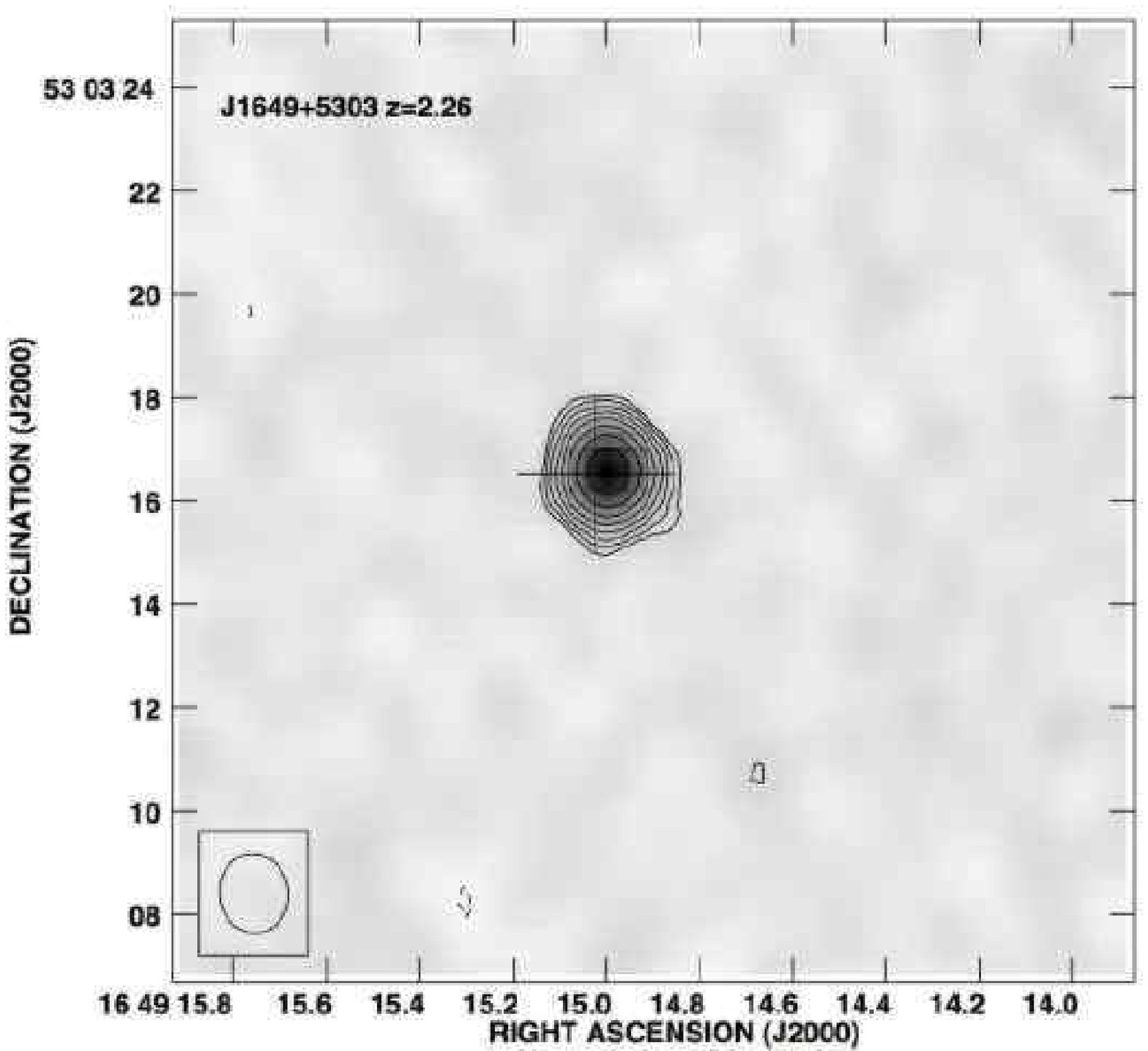}}}
\figurenum{1}
\caption{continued}
\end{figure}
\clearpage\newpage
\begin{figure}
{{\plottwo{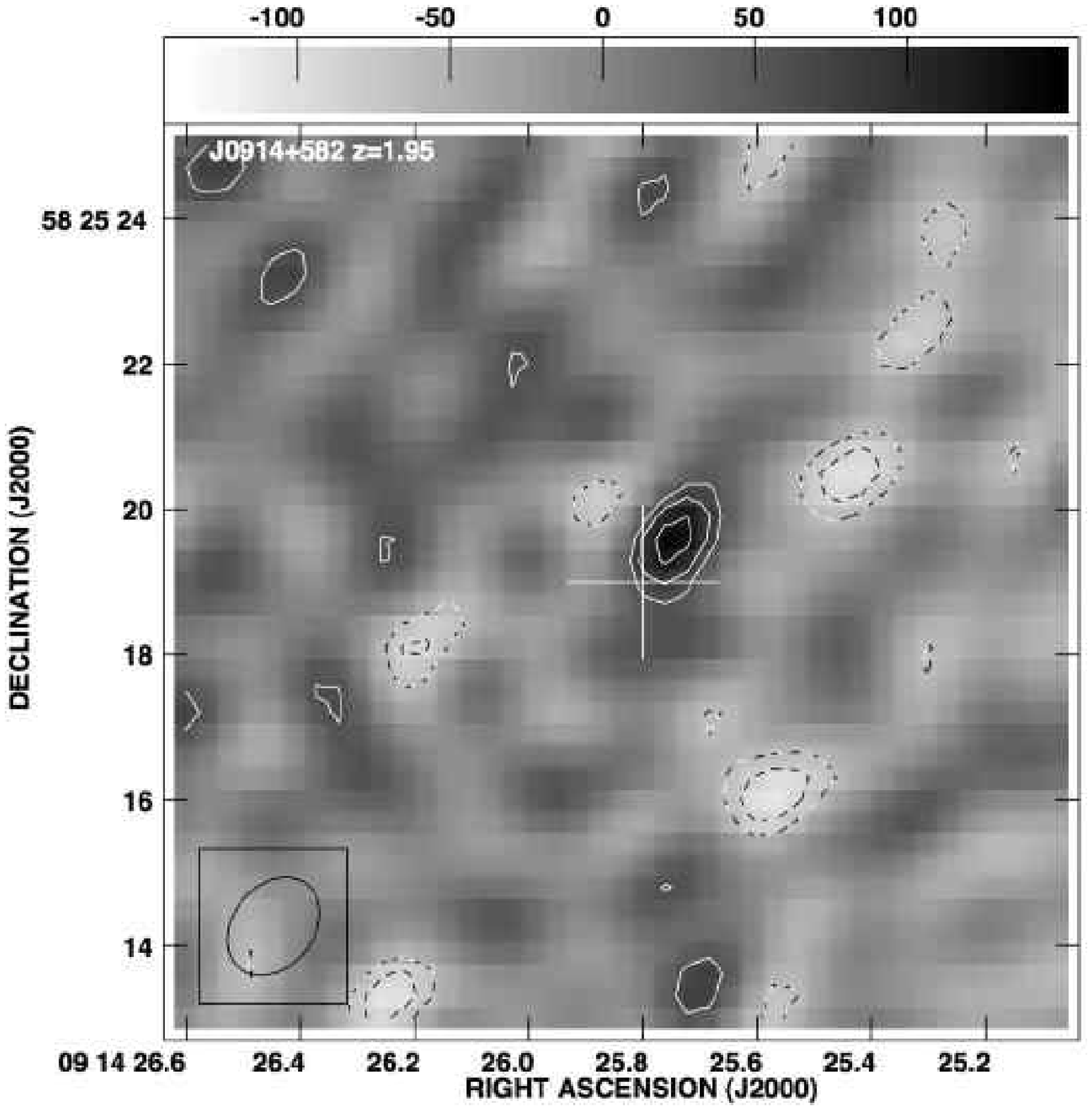}{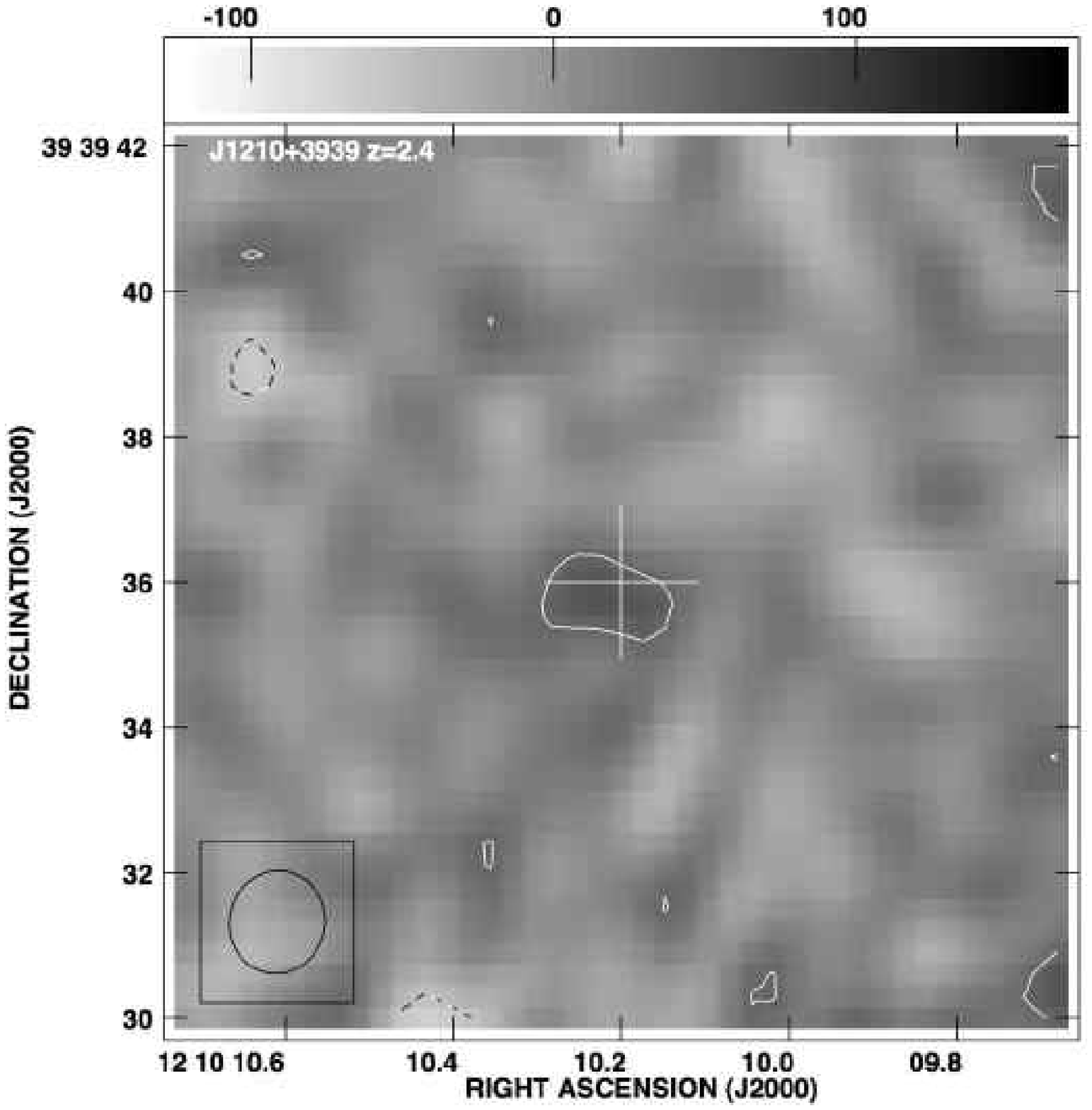}}
 {\plotone{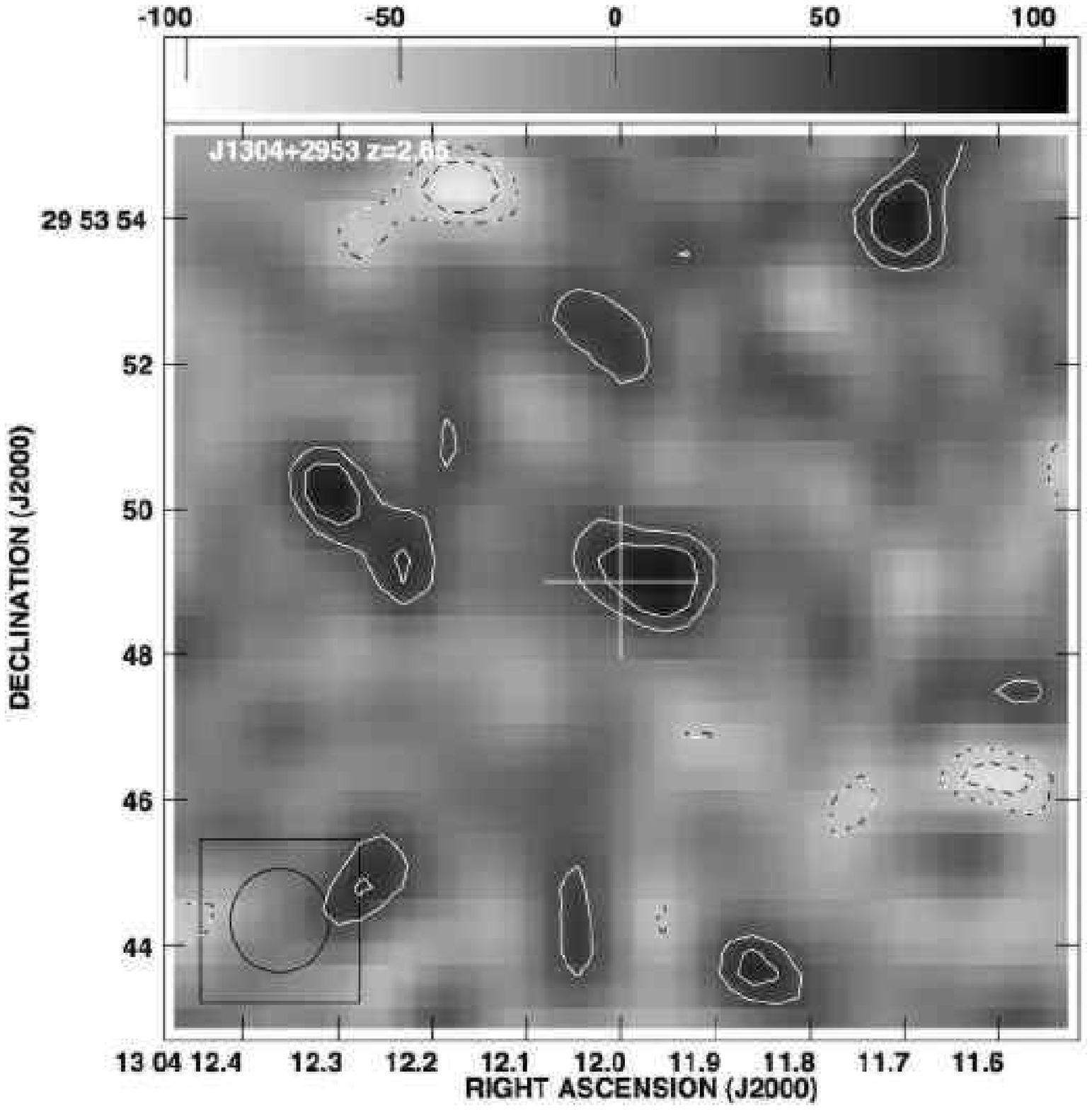}} 
}
\figurenum{1}
\caption{continued} 
\end{figure}
\pagestyle{empty} 
\clearpage\newpage
\begin{figure}[ht]
{{\plottwo{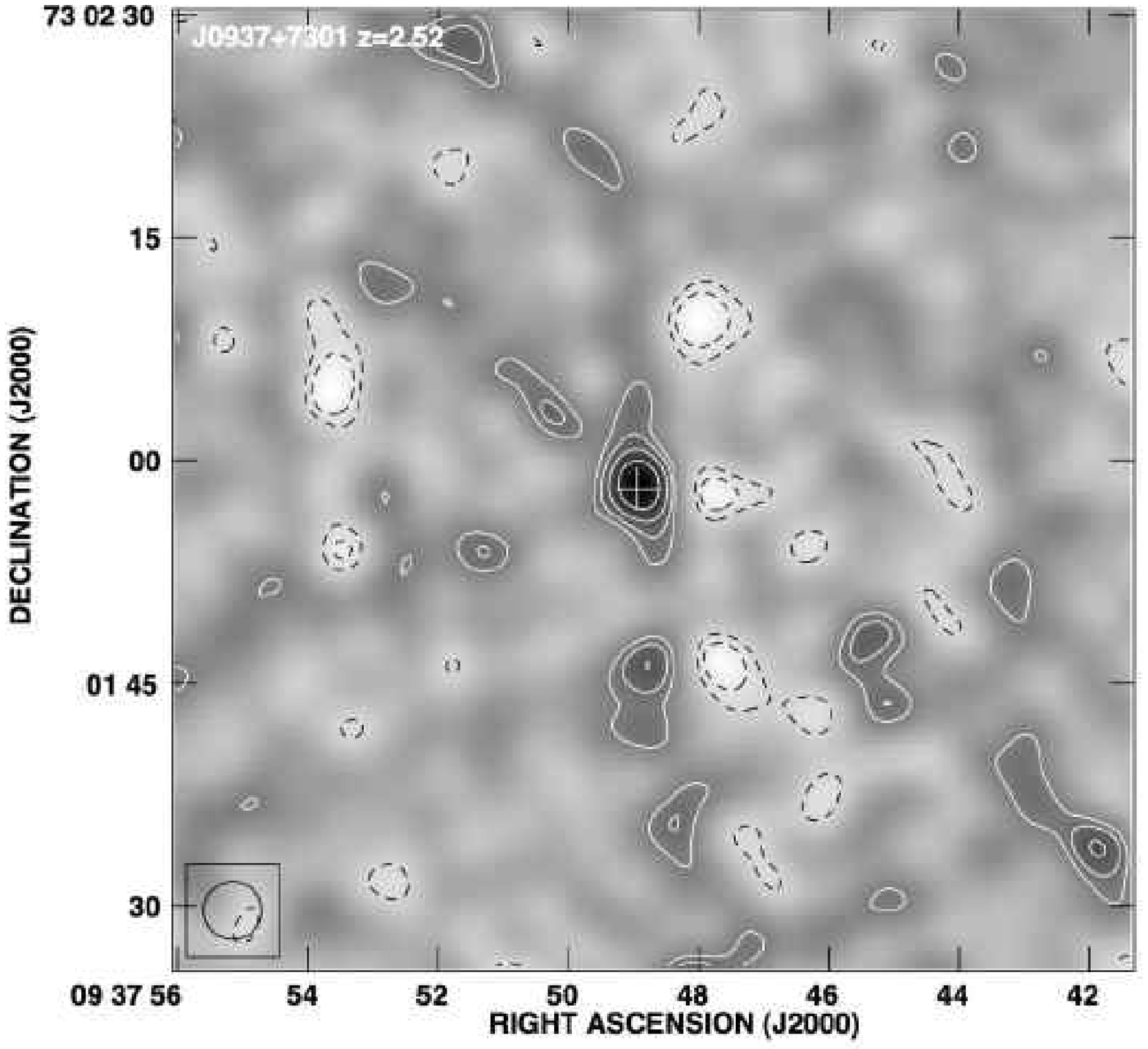}{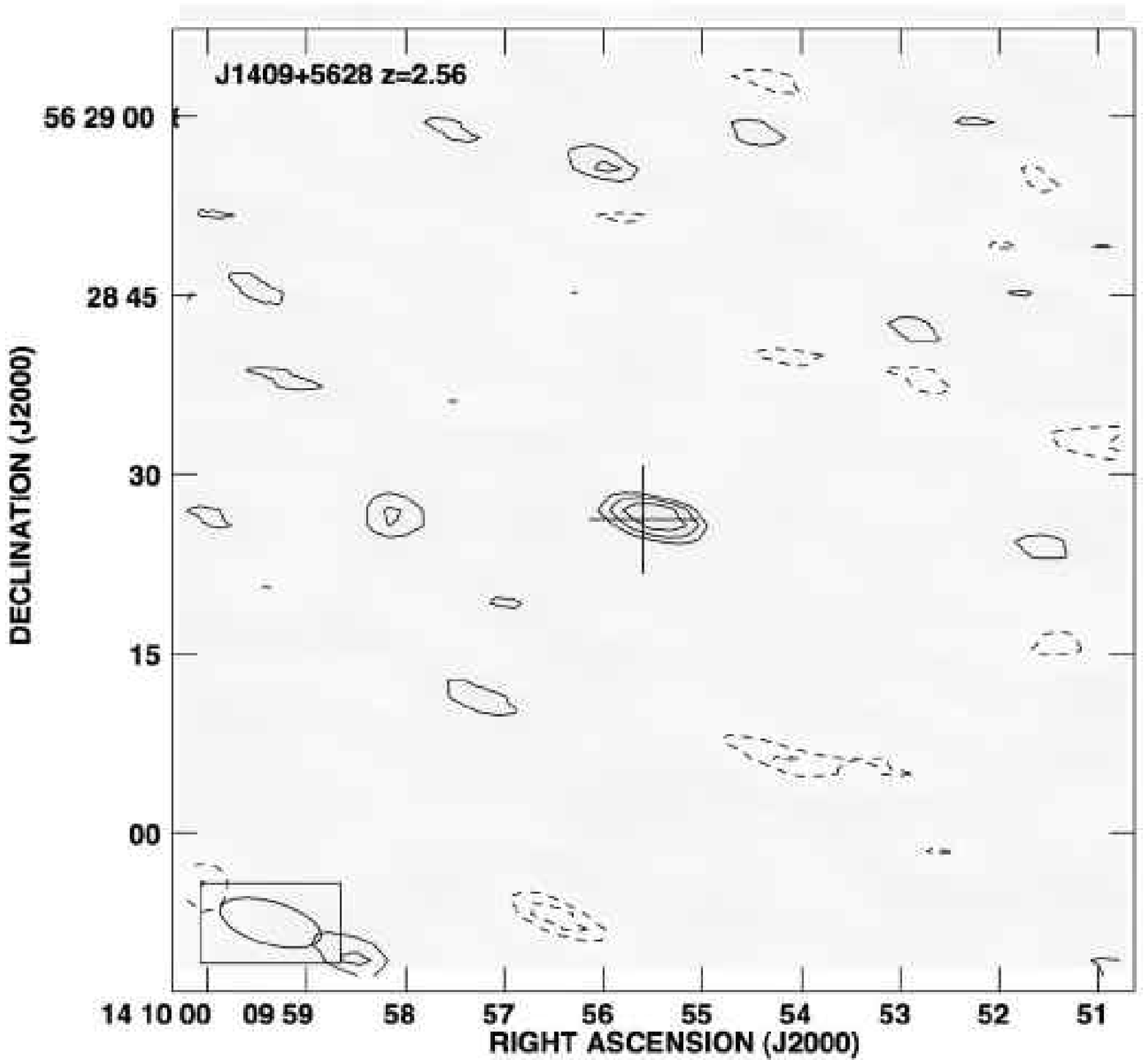}}
 {\plottwo{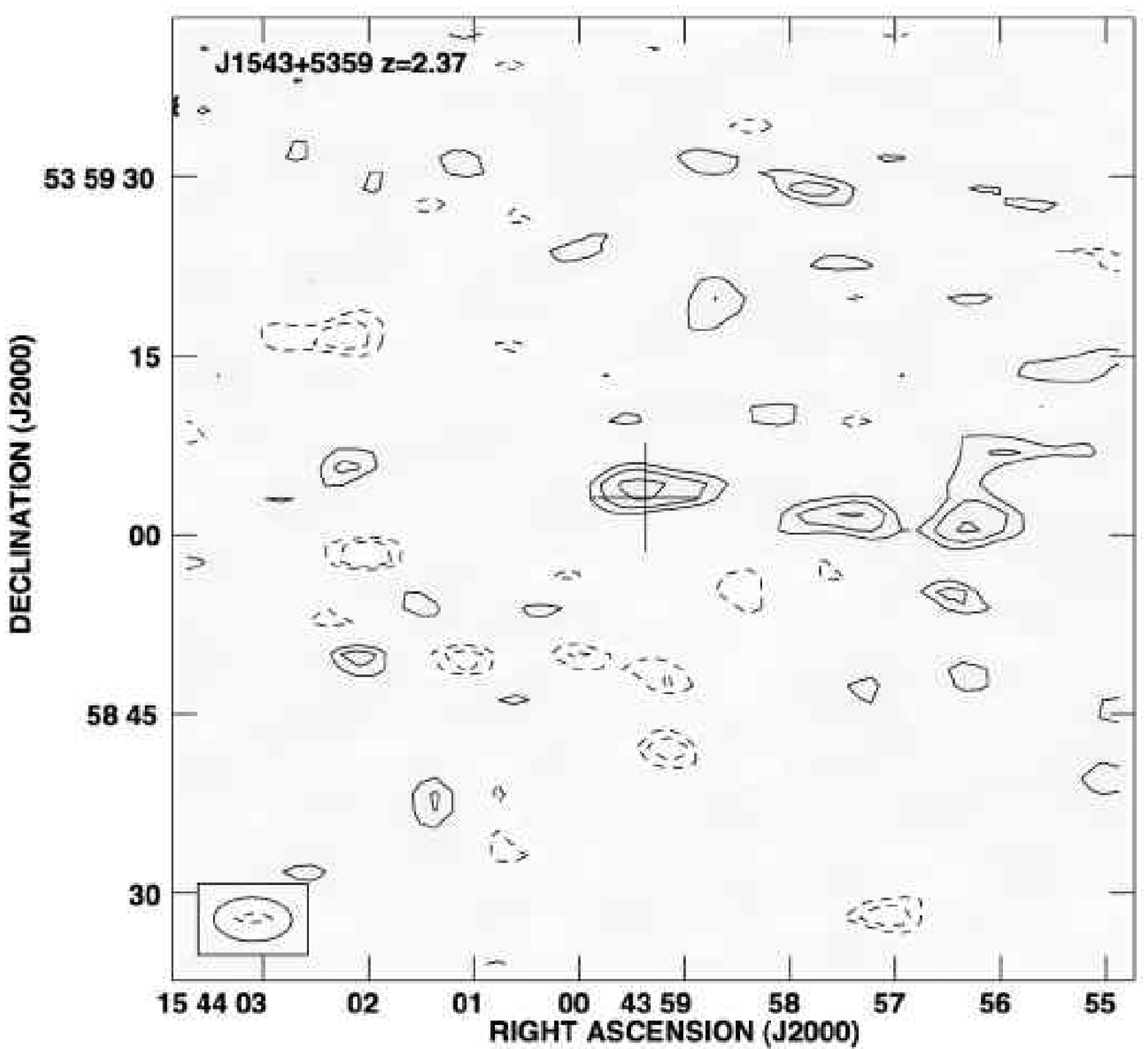}{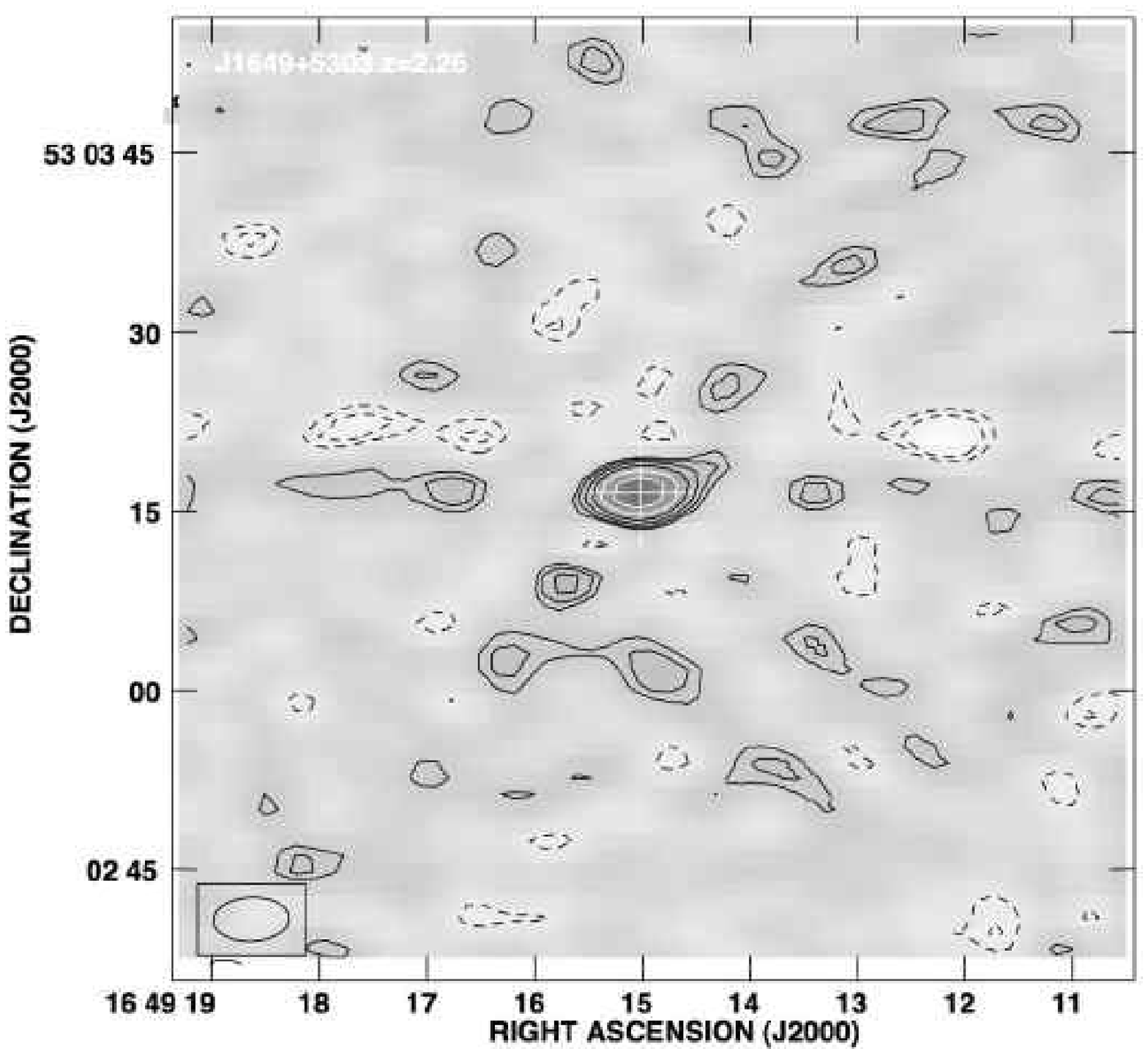}}}
\caption{Images at 5 GHz}
\end{figure}

\begin{figure}
{\plotone{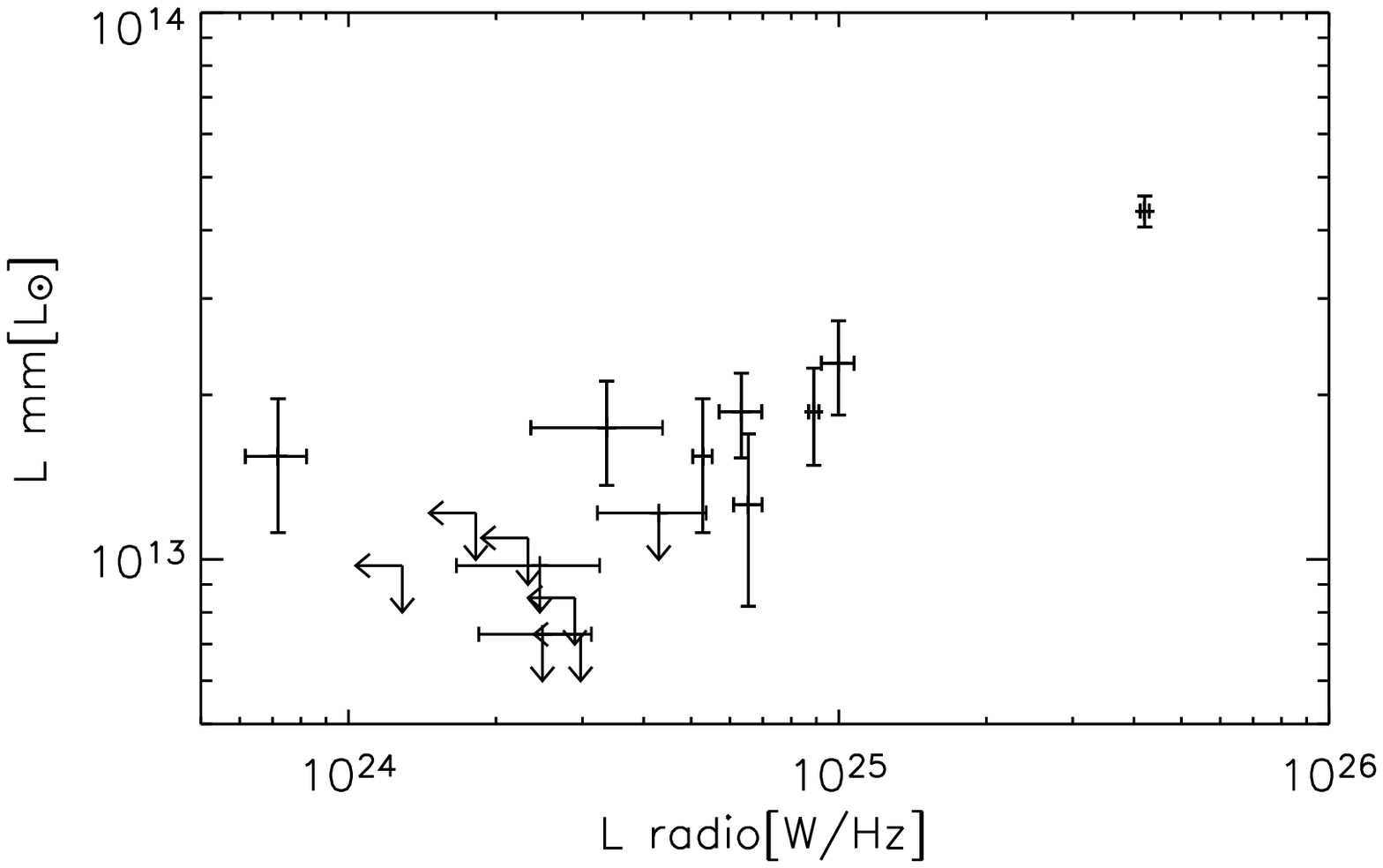}}
\end{figure}

\begin{figure} 
{\plotone{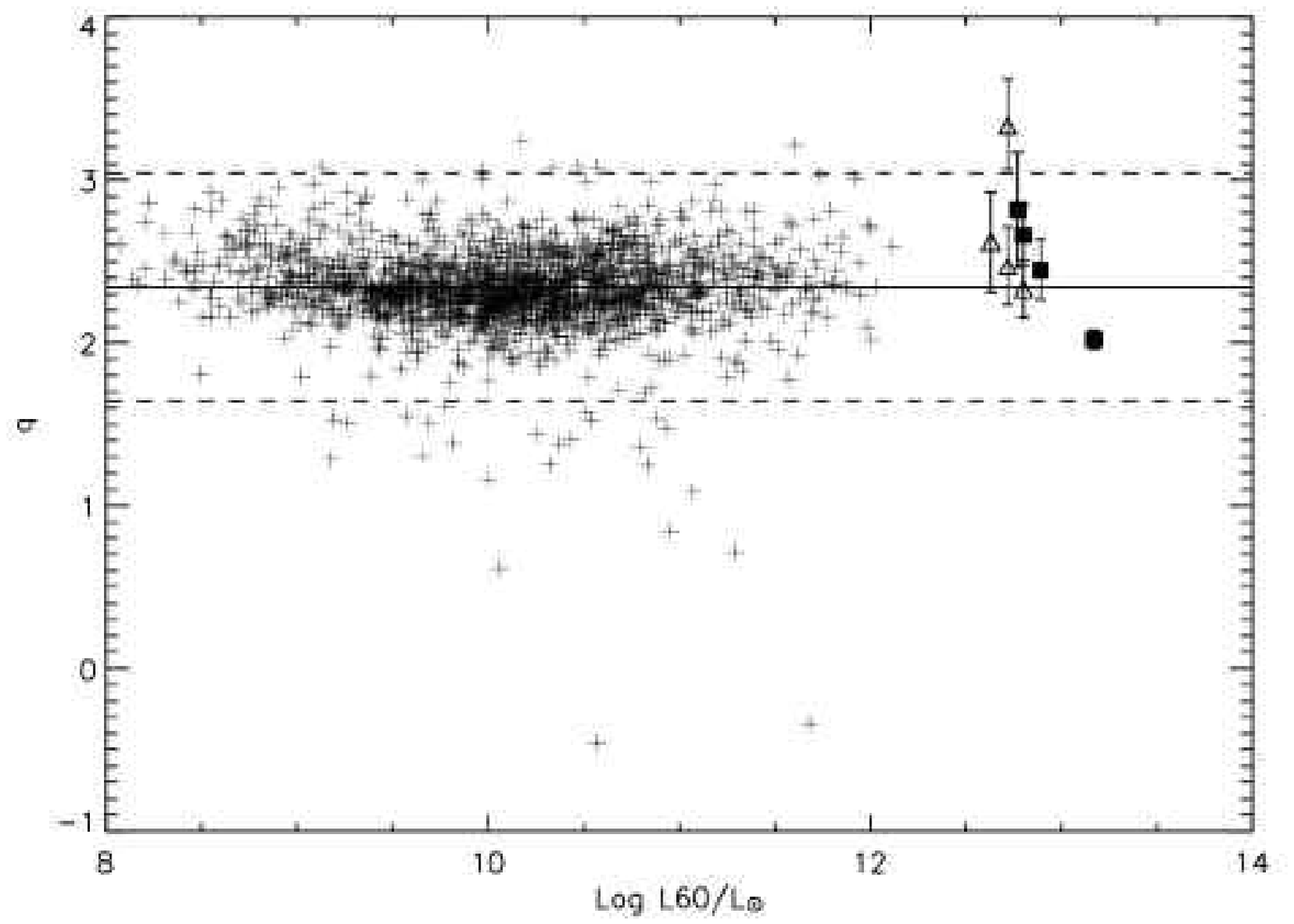}}
\end{figure}


\clearpage\newpage

\begin{references}
\reference{} Alexander, D.M., 2004, astroph-0401129; in {\it{Multiwavelength Mapping of Galaxy 
Formation and Evolution}}, R. Bender \& A. Renzini (eds), ESOUSM/MPE workshop-proceedings
\reference{} Alexander, D.M., et al. 2005a, Nature, 434, 738
\reference{} Alexander, D.M., et al. 2005b, ApJ, 632, 736
\reference{} Andreani, P., Cristiani, S., Grazian, A., La Franca, F., \& Goldschmidt, P. 2003, AJ 125, 444
\reference{} Appleton, P.N., Fadda, D.T., Marleau, F.R., Frayer, D.T. et al. 2004, ApJS, 154,147
\reference{} Becker, R. H., White, R. L., \& Helfand, D. J. 1995, ApJ, 450, 559
\reference{} Beelen,A., Cox,P., Benford, D.J.,, Dowell,C.D., et al. 2006, ApJ, in press: astroph0603121
\reference{} Benford, D.J., Cox, P., Omont, A., Phillips, T. G., \& McMahon, R. G.,1999, ApJ, 518, L65
\reference{} Bertoldi, F., Carilli, C. L., Cox, P., Fan, X., et al. 2003a, A\&A, 406, 55
\reference{} Bertoldi, F., Cox, P., Neri, R., Carilli., C.L., et al. 2003b, A\&A., 409, 4
\reference{} Beelen, A., Cox, P., Pety, J., Carilli, C.L., et al. 2004, A\&A, 423, 441
\reference{} Bridle, A.H., \& Schwab, F.R. 1999, in ASP Conf. Ser. 180, Bandwidth and 
Time-Average Smearing in Synthesis Imaging in Radio Astronomy II, ed. G.B.Taylor, C.L. Carilli,
 \& R. A. Perley (San Francisco: ASP), 357
\reference{} Carilli, C.L., Bertoldi, F., Omont, A., Cox, P. et
al. 2001a, AJ, 122, 1679 
\reference{} Carilli, C.L., Bertoldi, F., Rupen, M.P., Fan, X., et
al. 2001b, ApJ, 555, 625 
\reference{} Carilli, C.L., Cox, P., Bertoldi, F., Menten, K.M, et
al. 2002, ApJ, 575, 145
\reference{} Carilli, C.L., \& Yun, M.S., 1999, ApJ, 513, 13
\reference{} Carilli, C.L., Yun, M.S., 2000, ApJ, 530, 618
\reference{} Chapman, S.C., Blain, A. W., Smail, I, Ivison, R. J. 2005, ApJ, 622, 772
\reference{} Chini, R., Kreysa, E., Biermann, P.L., 1989, A\&A, 219, 87
\reference{} Condon, J.J. 1992, ARA\&A, 30, 575
\reference{} Condon, J.J. \& Yin, Z.F. 1990, ApJ 357, 97
\reference{} Cotton, W. D. 1999, in ASP Conf. Ser. 180, Synthesis Imaging in Radio Astronomy II, 
ed. G.B.Taylor, C.L. Carilli, \& R. A. Perley (San Francisco: ASP), 357
\reference{} Cox, P., Omont, A., Djorgovsky, S.G., Bertoldi, F., et al. 2002, A\&A 387, 406
\reference{} Deutsch, E.W. 1999, AJ, 118, 1882
\reference{} Elbaz, D., Cesarsky, C. J., Chanial, P., Aussel, H., et al. 2002, A\&A 284, 848
\reference{} Elvis, M., Wilkes, B. J., McDowel, J.C., Green, R.F.,. et al. 1994, ApJS, 95, 1 
\reference{} Feigelson, E. D., \& Nelson, P. I.1985, ApJ, 293, 192
\reference{} Fomalont, E.B., 1999, in ASP Conf. Ser. 180, Bandwidth and 
Time-Average Smearing in Synthesis Imaging in Radio Astronomy II, ed. G.B.Taylor, C.L. Carilli,
 \& R. A. Perley (San Francisco: ASP), 35
\reference{} Ferrarese, L., \& Merritt, D. 2000, ApJ, 539, L9
\reference{} Gebhardt, K., Kormendy, J., Ho, L., Bender, R., et
al. 2000, ApJ, 543, L5
\reference{} Greve, T.R., Ivison, R.J., Bertoldi, F., Stevens, J.A., et al. 2004, MNRAS 354, 779
\reference{} Haas, M., , Muller, S.A.H., Chini., R., Meisenheimer, K., et al. 2000, A\&A 354, 45
\reference{} Haas, M., Klaas, U., M\"uller, S.,A., Bertoldi, F., et al. 2003, A\&A 402, 87
\reference{} Hopkins, P.F., Hernquist, L. Cox, L., et al. 2005, submitted to ApJ, astroph/05036398
\reference{} Isaak, K.G., Priddey, R.S., McMahon, R.G., Omont., A., et al., 2002, MNRAS, 329, 149I
\reference{} Isobe, T., Feigelson, E.D., \& Nelson, P. I., 1986, ApJ, 306, 490
\reference{} Kreysa, E., et al. 1998, Proc. SPIE, 3357, 319
\reference{} Magorrian, J., Tremaine, S., Richstone, D., Bender, R., et al. 1998, AJ, 115, 2285
\reference{} McMahon, R.G., Omont, A., Bergeron, J., Kreysa, E., \& Haslam, C.G.T., 1994, MNRAS, 267
\reference{} Miller, N.A., \& Owen, F.N., 2001, AJ, 121	
\reference{} Momjian, E., Petric, A., \& Carilli, C. L., 2004, AJ, 127, 587
\reference{} Momjian, E., Carilli, C.L., \& Petric, A.O., 2005, AJ 129, 1809
\reference{} O'Dea, C.P. 1998, PASP, 110, 493
\reference{} Omont, A., Beelen, A., Bertoldi, F. et al. 2003, A\&A, 398, 857
\reference{} Omont, A., Cox., P., Bertoldi, F., McMahon, R.G., 2001, A\&A., 374, 371
\reference{} Page, M.J., 2001, MNRAS, 325, 575
\reference{} Page, M.J., Stevens, J.A., Mittaz, J.P.D., Carrera, F.J., 2001, Science 294,  2516
\reference{} Page, M.J., Stevens, J.A., Ivison, R.J., Carrera, F.J., 2004,ApJ, 611, L85
\reference{} Petric, A.O., Carilli, C.L., Bertoldi, F., Fan, X., et al. 2003, AJ 126, 15
\reference{} Priddey, R. S., Isaak, K. G., McMahon, R. G., \& Omont, A., 2003, MNRAS, 339, 1183
\reference{} Springel, V., Di Matteo, T., \& Hernquist, L. 2005, ApJ 620, 79
\reference{} Sopp, H. \& Alexander, P., 1991, MNRAS, 251P, 14
\reference{} Tremaine, S., Gebhardt, K., Bender, R., Bower, G., et al. 2002, ApJ, 574, 740 
\reference{} Walter, F., Bertoldi, F., Carilli, C.L., Cox, P., et al. 2003, Nature, 424, 406
\reference{} Yun, M.S., Reddy, N.A., \& Condon, J.J., 2001, ApJ, 554, 803
\reference{} Yun,M. S., \& Carilli, C. L., 2002, ApJ, 568, 88
\end{references}
\end{document}